--------

\documentstyle[pra,aps,preprint,eqsecnum]{revtex}
\newcommand{\la}{\langle}
\newcommand{\ra}{\rangle}
\newcommand{\un}{\underline}
\begin{document}
\draft
\preprint{\vbox{\baselineskip=12pt
\rightline{IMPERIAL/TP/95-96/44}
\rightline{}}}
\title{\bf{Decoherence and Localization in Quantum Two-Level  Systems}} 
\author{Ting Yu\footnote{Electronic address: ting.yu@ic.ac.uk}}
\address{Theoretical Physics Group, Blackett Laboratory, Imperial
College,\\
 Prince Consort Road, London SW7 2BZ, United Kingdom}

\date{May 26, 1996}
\maketitle

\begin{abstract}
We study and compare the decoherent histories approach, the 
environment-induced decoherence and the localization properties
of the solutions to the stochastic Schr\"{o}dinger equation 
in quantum jump simulation  and 
quantum state diffusion approaches, for a quantum two-level 
system model. 
We show, in particular, that there is a close connection 
between the decoherent histories and the quantum jump simulation,
complementing a connection with the quantum state diffusion 
approach noted earlier by Di\'{o}si, Gisin, Halliwell and 
Percival. In the case of the decoherent histories 
analysis, the degree of approximate decoherence is discussed 
in detail.
\end{abstract}
\newpage
    
\section{INTRODUCTION}
\label{sec:intro}
 
Two primary paradigms -- the environment-induced decoherence approach,  
proposed by Zurek \cite{zur,zur1,zur2},  and  the consistent histories
approach by Griffiths \cite{gri} and later  by Omn\`{e}s \cite{omn} 
and by Gell-Mann and Hartle \cite{gel,gel1,har} have been recently 
developed to solve the fundamental issues in quantum theory, 
especially, quantum measurement problems and the transition from quantum
to classical. The environment-induced decoherence emphasizes the
division between the system and its environment. The interaction of the
system with its environment is responsible for  the decay of the 
quantum coherence of the system. The decoherent histories 
approach is designed to provide the most general descriptions for a 
closed system by using the concept of history -- a sequence of events 
at a succession of times. Both approaches are applicable to  open 
quantum systems. 

Another set of viable theories within the framework of quantum 
mechanics are
the various unravellings of master equation
as stochastic Schr\"{o}dinger equations for the single member of the
ensemble.  Among others, the quantum state diffusion and
the quantum jump simulation approaches have been extensively studied 
in recent years ({\it e.g.}, see \cite{gis,car,gar,mil}). As
phenomenological theories, these stochastic approaches are not only of 
theoretical interest but also of practical value.

The open quantum system provides a unified framework  to
exhibit the properties of the various approaches we have mentioned 
above. The master equation which describes the evolution of the open 
quantum system plays a central role in the investigations into the decay of 
quantum coherence due to the interaction with a much larger 
environment. However, it does not tell us how an individual member 
of an ensemble evolves in a dissipative environment. The unravelling
of  master equation as stochastic Schr\"{o}dinger equation could provide 
such a description within its domain of applicability. 
Corresponding to the decoherence process in the density operator
formalism, in stochastic Schr\"{o}dinger equation approaches, 
the solution to the stochastic Schr\"{o}dinger equation often possesses 
a very remarkable property -- the solution tends to  localize at
 some special states after a localization time
scale. For quantum diffusion approach, this localization property 
has been justified in many different 
situations \cite{dio,per,gis1,gar1,dio1,hal}. 

It is worth emphasizing  that a key point in these approaches is 
the mutual influence between the system of interests and its 
environment. This mutual influence is the common sources of the 
many different phenomena such as dissipation, fluctuation,
decoherence, localization, {\it etc}. 
     
Analysis of decoherence and  localization properties is usually rather
involved. The entanglement of the complicated mathematics and the subtle 
conceptual issues  often tends to make the detail scrutiny of the basic concepts
impossible. The attractiveness of two-level system model is that,
perhaps it is   one of the simplest yet physical meaningful models.

The purpose of this  paper is to employ a widely used two-level 
system model as  a unified framework to examine the dynamics of the
open quantum system by the decoherent histories, the
environment-induced decoherence and the stochastic Schr\"{o}dinger 
equations. We mainly consider the decoherence process and the localization
process. The various time scales concerning
these processes  are discussed. We provide a detailed analysis of the
degree of decoherence
which is important for a real physical process.

The plan of this paper is as follows. In Section II,  we briefly
present our two-level system model and its basic properties. 
We study the consistent histories approach, the degree of decoherence in 
Sections III and IV,  respectively.  We study the unravelling of master
equation and the localization properties of the solutions to the stochastic
Schr\"{o}dinger equations in both quantum jump simulations and quantum
state diffusions  in Section V.  We make some
remarks in Section VI. In  Appendix we present a proof of Theorem in
Section IV.

\newpage
\section{ The Model}
\label{sec:two}
A fundamental building block in quantum theory is the two-level system.  
We consider a two-level atom system, which is radiatively damped
by its interaction with the many modes of a radiation field in
thermal equilibrium at temperature $T$. The upper level and lower
level are denoted by $|2\ra$ and $|1\ra$, respectively. A widely used 
 master equation for the two-level atom takes standard Lindblad 
form (in the Schr\"{o}dinger picture) ({\it e.g.},
see \cite{car,gar}):
\begin{eqnarray}
\label{master}
\dot \rho  = &-& {i\over\hbar} [H, \rho ]\nonumber \\
             &+& {\gamma\over 2}(\overline {n}+1) (2a\rho a^{\dag} -
 a^{\dag} a\rho -\rho a^{\dag} a)\nonumber \\
             &+& {\gamma \over 2}\overline {n}(2a^{\dag} \rho a -
aa^{\dag} \rho - \rho aa^{\dag})\>. 
\end{eqnarray}
Here, the Hamiltonian of
the atom in the absence of the environment is given by 
\begin{equation}
H={\hbar\omega\over 2}\sigma_z.
\end{equation}
The Lindblad operators,  which model the effects of the environment in 
this situation, are
\begin{equation}
L_1=\sqrt{\gamma(\overline{n}+1)}a, \> L_2=\sqrt{\gamma\overline{n}}a^{\dag}.
\end{equation} 
The transition rate from $|2\ra \rightarrow |1\ra$ is described by the term
proportional to  $({\gamma /2})(\bar n + 1)$, and the transition
rate from $|1\ra \rightarrow |2\ra$ is described by the term proportional
to $({\gamma /2})\bar n$. We use $\sigma_x, \sigma_y $ and $\sigma_z$
to denote  Pauli matrices  and 
$a, a^{\dag}$  atomic lowering and raising operators, which are defined in
the usual way
\begin{equation}
\sigma_x=\left[\begin{array}{cc}
                 0 & \, 1 \\
                 1 & \, 0 
                \end{array}\right],\>\> 
                \sigma_y=\left[\begin{array}{cc}
                 0 &  -i \\
                 i &  0 
                \end{array}\right],\>\>
                \sigma_z=\left[\begin{array}{cc}
                 1 &  0 \\
                 0 &  -1 
                \end{array}\right]
\end{equation}
and 
\begin{equation}
a={1\over 2}(\sigma_x-i\sigma_y), \>\> a^{\dag}={1\over 2}(\sigma_x+i\sigma_y)\>.
\end{equation}
The master equation (\ref{master}) in the basis $|2\ra, |1\ra$ may be written as
\begin{eqnarray}
\label{solution}
\dot \rho_{11}&=&-\gamma(\overline
                {n}+1)\rho_{11}+\gamma\overline {n}\rho_{22}\>,\\
\dot\rho_{22} &=& \gamma(\overline {n}+1)\rho_{11}-
\gamma\overline {n}\rho_{22}\>,\label{da0}\\
\dot\rho_{12} &=&-[{\gamma\over 2}(2\overline {n} +1)
 +i\omega]\rho_{12}\>,\label{da1}\\
\dot\rho_{21} &=&-[{\gamma\over 2}(2\overline {n}+1) 
-i\omega]\rho_{21}\>.\label{da2}
\end{eqnarray}
The  general solutions to Eqs.
(\ref{solution})-(\ref{da2}) are as follows
\begin{eqnarray}
\rho_{11}(t)&=&{\overline{n}\over
               \overline{n}+1}B_1+ B_2e^{-\gamma(2\overline n +1)t}\>,\label{so}\\
\rho_{22}(t)&=& B_1 - B_2e^{-\gamma(2\overline n +1)t}\>,\label{so1}\\
\rho_{12}(t)&=&B_3e^{-[{\gamma\over 2}(2\overline n +1)+i\omega] t}\>,\label{so2}\\
\rho_{21}(t)&=&B_4e^{-[{\gamma\over 2}(2\overline n +1)-i\omega]t}\>,\label{so3}
\end{eqnarray}
where $B_i\,(i=1,2,3,4)$ are arbitrary constants which can be easily
determined once the initial condition is given. For the initial
density
matrix with ${\rm Tr}(\rho_0)=1$ we easily get
\begin{eqnarray}
B_1&=& {{\overline{n}+1}\over 2\overline{n}+1}\\
B_2&=& \rho_{11}(0)-{\overline{n}\over 2\overline{n}+1}\\
B_3&=&\rho_{12}(0)\\
B_4&=&\rho_{21}(0)
\end{eqnarray}

In the basis $|2\ra,|1\ra$ the off-diagonal elements tend to zero as $t$
goes to infinite. The density operator $\rho$ tends to the stationary
density operator $\rho_s$:
\begin{equation}
\rho\longrightarrow \rho_s =\left[\begin{array}{cc}
                        {\overline {n}\over {2\overline {n}} + 1} & \, 0 \\
                         0& \, {\overline {n}+1\over {2\overline {n}} + 1}
                         \end{array}\right].
\end{equation}
This is an elementary example of   environment-induced decoherence.
Diagonalization occurs in the basis $|2\ra,|1\ra$.  From
equations (\ref{da1}) and (\ref{da2}), it follows that the decoherence
time scale
$t_D$ is given by
\begin{equation}
\label{scale}
t_D\sim {1\over \gamma(2\overline{n}+1)}\>.
\end{equation}
  
It is easily shown that master equation (\ref{master}) is invariant under
unitary transformations of the Lindblad operators:
\begin{equation}
a\longmapsto UaU^{\dag},\>\>a^{\dag}\longmapsto  Ua^{\dag}U^{\dag},
\end{equation}
where $U$ is a unitary matrix. Correspondingly, the density operator
$\rho$ transforms in the same way:
\begin{equation}
\rho \longmapsto U\rho U^{\dag}.
\end{equation}
Thus,  when $t\rightarrow \infty$
\begin{equation}
\rho\longrightarrow U\rho_sU^{\dag}.
\end{equation}
Generally, the density matrix $U\rho_sU^{\dag}$ is no longer
diagonal. This indicates that environment-induced decoherence does not
occur in other bases.

\section{Consistent histories in the two-level system model}
\label{sec:consist}

The decoherent histories approach \cite{gri,omn,gel,gel1,har} offers
a sensible way to assign probabilities to a sequence of properties of
a quantum system without referring to the measurements or to a 
classical domain. 

A history is defined in general as a sequence of properties of a
closed system occurring at different times, which  is
denoted as
\begin{equation}
C_{\alpha}= P^{(n)}_{\alpha_n}(t_n), ... , P^{(1)}_{\alpha_1}(t_1),
\end{equation}
where $P^{(i)}_{\alpha_i}(t_i)$ are the projection
operators in the Heisenberg picture at times $t_i$:
\begin{equation}
P^{(i)}_{\alpha_i}(t_i)=e^{{i\over \hbar}(t_i-t_0)H}
P^{(i)}_{\alpha_i}e^{-{i\over\hbar}(t_i-t_0)H},
\end{equation}
here, $H$ is the Hamiltonian of the closed system. These projection
operators satisfy 
exhaustive and exclusive conditions:
\begin{equation}
\sum_{\alpha_i}P^{(i)}_{\alpha_i}(t_i)=I, \>
\>\>P^{(i)}_{\alpha_i}(t_i)P^{(i)}_{\beta_i}(t_i)=
\delta_{\alpha_i\beta_i}P^{(i)}_{\alpha_i}(t_i).
\end{equation}
The superscript $(i)$ labels the set of projections used at time $t_i$  and $\alpha_i$
denotes the particular alternative.

A natural way to assign the probability to a history is
\begin{eqnarray}
\label{prohis}
p(C_{\alpha})& = & {\rm Tr}(C_{\alpha}\rho(t_0)C^{\dag}_{\alpha})\nonumber\\ 
             & = & {\rm Tr} \left(P^{(n)}_{\alpha_n}(t_n)...
P^{(1)}_{\alpha_1}(t_1)\rho(t_0)P^{(1)}_{\alpha_1}(t_1)... P^{(n)}_{\alpha_n}(t_n)\right).
\end{eqnarray}      
However, one finds that (\ref{prohis}) generally does not satisfy the
usual probability sum rules. The necessary and sufficient condition to
 guarantee that the probability sum rules hold  is that
the real part of the decoherence functional $D(\alpha,\alpha')$ vanishes,
\begin{equation}
\label{mad}
{\rm Re}D(\un{\alpha},\un{\alpha}')={\rm Re}\,{\rm
Tr}(C_{\alpha}\rho(t_0)C^{\dag}_{\alpha'})=0
\end{equation}
for any two different histories $C_{\alpha}$ and $C_{\alpha'}$.
The sets of histories
satisfying (\ref{mad}) are said to be {\it consistent} (or {\it weakly
decoherent}). 
Physical mechanisms causing (\ref{mad}) to be satisfied typically
lead also to the stronger condition 
\begin{equation}
\label{mad1}
D(\un{\alpha},\un{\alpha}')=0, \>\> \forall  \un{\alpha}\neq \un{\alpha}'
\end{equation}
which is called {\it medium decoherence}\cite{gel1}. (In this paper,
we  simply refer it as decoherence)

Although the decoherent histories approach was primary 
designed for a closed system, the approach is of particular 
importance for the open system which may be regarded as  a
subsystem of a large closed system. For the open quantum system,
a natural coarse-graining is to focus only on the properties 
of the distinguished system whilst ignoring the environment. 
In this case, a natural selection of projections at each
time  is of form $P_{\alpha}\otimes I^{\mathcal E} $, where
$P_{\alpha}$ 
is a projection onto the distinguished subsystem and 
$I^{\mathcal E}$ denotes the identity projection on the
environment. In the  Markovian regime, the decoherence 
functional could be constructed entirely in terms of the
reduced density matrix of the system \cite{paz}:
\begin{equation}
\label{decoherence}
D[\underline\alpha,\underline\alpha']={\rm
Tr}\left(P^{(n)}_{\alpha_n}K^{t_n}_{t_{(n-1)}}[P^{(n-1)}_{\alpha_{n-1}}
...
K^{t_2}_{t_1}[P^{(1)}_{\alpha_1}K^{t_1}_{t_0}[\rho_0]P^{(1)}_{\alpha_1'}]
... P^{(n-1)}_{\alpha_{n-1}'}]P^{(n)}_{\alpha_n'}\right),
\end{equation}
where the trace is taken over the distinguished system only. 
The quantity $K^{t_k}_{t_{k-1}}[\>\cdot \>]$ is the reduced 
density operator propagator: $\rho_t=K^t_0[\rho_0]$. 

In what follows we shall make a detailed analysis of the 
decoherent histories in the two-level model described by 
the master equation (\ref{master}) which depicts a
 Markovian process. 

First, let us consider the projection operators represented by
\begin{equation}
\label{3.3}
P_1=|1\ra\la1| \> \> {\rm and}\>\> P_2=|2\ra\la2|\>.
\end{equation}
Obviously, $\{P_i, i=1,2\}$ form a set of complete and exclusive
projection operators. Physically,  $P_1$ may represent that the atom emits
a photon whereas  $P_2$ may represent that the atom absorbs a photon.
Then the decoherence functional at  two time
points is given by
\begin{equation}
\label{func}
D[\underline\alpha,\underline\alpha']=
\delta_{\alpha_2 \alpha_{2'}}{\rm Tr}\left
(P^2_{\alpha_2}K^{t_2}_{t_1}\left [P^1_{\alpha_1} K^{t_1}_{0}\left
[\rho_0\right ]P^1_{\alpha_1'}\right ]\right ).
\end{equation}
It is easily shown that, for any $2\times 2$  matrix $A$, the 
matrix $ P_iAP_j\, (i\neq j)$ is an upper (or a lower) triangle
matrix. From equations (\ref{da1}),(\ref{da2}) we know that 
$K^{t_{i+1}}_{t_i}[\>\cdot\>]$ propagates the matrix with zero
diagonal elements into the matrix with zero diagonal
elements.  So, for any initial density matrix $\rho_0$,  the trace in
Eq. (\ref{func}) is exactly zero for
any different pairs of histories $(\un{\alpha} \neq \un{\alpha}')$ and
for any interval $t_2-t_1$. This
demonstrates that the set of histories consisting of 
projectors (\ref{3.3})  are exactly decoherent. The generalization 
to $n$ time points is straightforward. The exact decoherence for any
time interval is slightly surprising. (The density matrix, by
contrast, only becomes exactly diagonal as $t \rightarrow \infty$).  
This exactness is due to the simplicity of the model and we do not 
expect it to be a generic feature.

Next, consider more general projection operators which correspond to
the projection to any direction.
 With  any direction denoted by a unit vector
 $n =(\sin\theta\cos\phi, \sin\theta\sin\phi, \cos\phi)$, we associate a
vector $| n\ra$ which belongs to the
Hilbert space of the two-level system,

\begin{equation}
|n\ra =\cos{\theta\over 2}|1\ra -e^{-i\phi}\sin{\theta\over 2}|2\ra\>.
\end{equation}
Then one  can define the following
projection operators on the Hilbert space of the  system:
\begin{equation}
P_+=|n\ra\la n|,\>\>\> P_-=|n'\ra\la n'|\>,
\end{equation}
where $| n'\ra$ is the orthogonal complimentary of $|n\ra$:
\begin{equation}
| n'\ra = e^{i\phi}\sin{\theta\over 2}|1\ra+\cos{\theta\over 2}|2\ra\>.
\end{equation}

We shall show that a set of histories 
consisting of the projection operators $P_+$ and $P_-$ are 
approximately decoherent. To this end,
first, note that for any $2\times 2$ matrix $A$,
\begin{equation}
 {\rm Tr}\,( P_+AP_- )= {\rm Tr}\,( P_-AP_+) =0\>. 
\end{equation}
Hence, from (\ref{so})--(\ref{so3}), it can be seen that, after
the propagation of
$K^{t_{i+1}}_{t_i}[\>\cdot\>]$,  
all of the diagonal elements of
matrix $K^{t_2}_{t_1}[P_{\pm}K^{t_1}_{t_0}[\rho_0]P_{\mp}]$ contain an
exponential damping factor
\begin{equation}
{\rm Damping\>\, factor}=e^{-\gamma(2{\overline n} +1)t}\>.
\end{equation}
Thus, we conclude
\begin{equation}
D[\underline\alpha,\underline\alpha']\approx 0, \> 
\forall \un{\alpha}\neq\un{\alpha}'\>.
\end{equation}
This  proves  that the set of histories consisting of  
$P_+, P_-$ are approximately
decoherent if time interval between $t_k$ and $t_{k+1}$ is larger than 
the characteristic time scale, 
\begin{equation}
\label{add}
t_{\rm decoherence}\sim {1\over\gamma ({2\overline {n} + 1})}.
\end{equation}
This is an expected result. We will give a more detailed estimate 
of the degree of decoherence in the next section. 

Note that $t_{\rm decoherence}$ decreases as
the coupling $\gamma$ is made stronger. To see the influence of the
temperature of the bath, a simple expression for $\overline {n}$ may
be chosen as 
\begin{equation}
\overline {n}={1\over e^{\hbar \omega /k_BT}-1}\>.
\end{equation}
From this, it is easy to see that the decoherence is more effective
if the temperature of bath, $T$, increases.  Conversely, decreasing
temperature will make system spend more time to  decohere. The maximum
decoherence time scale is $1/\gamma$ which corresponds to zero
temperature of bath. 

In summary, we find exact decoherence of histories characterized 
by the projections onto $|1\ra$ and $|2\ra$, and approximate 
decoherence in any other basis. 

Finally, we examine the probabilities for two times histories 
consisting of projections (\ref{3.3}). These are given
by
\begin{eqnarray}
p(1,2)&=&{\rm Tr}\left(P_1K^{t_2}_{t_1}
[P_2K^{t_1}_{t_0}[\rho_0]P_2]\right)\nonumber\\
      &=&{\overline{n}+1\over{2\overline{n}+1}}
(1-\delta)\rho_{11}(t_1),\\
p(2,1)&=&{\rm Tr}\left(P_2K^{t_2}_{t_1}
[P_1K^{t_1}_{t_0}[\rho_0]P_1]\right)\nonumber\\
      &=&{\overline{n}\over{2\overline{n}+1}}(1-\delta)\rho_{22}(t_1),\\
p(1,1)&=&{\rm Tr}\left(P_1K^{t_2}_{t_1}
[P_1K^{t_1}_{t_0}[\rho_0]P_1]\right)\nonumber\\
      &=&{1\over{2\overline{n}+1}}
\left[\overline{n}+(\overline{n}+1)\delta\right]\rho_{11}(t_1),\\
p(2,2)&=&{\rm Tr}\left(P_2K^{t_2}_{t_1}
[P_2K^{t_1}_{t_0}[\rho_0]P_2]\right)\nonumber\\
      &=&{1\over{2\overline{n}+1}}
\left[(\overline{n}+1)+\overline{n}\delta\right]\rho_{22}(t_1).
\end{eqnarray}
Where
\begin{eqnarray}
\rho_{11}(t_1)&=&{\overline{n}\over{2\overline{n}+1}}+
\left[\rho_{11}(0)-{\overline{n}\over
{2\overline{n}+1}}\right]\delta,\\
\rho_{22}(t_1)&=&{\overline{n}+1\over{2\overline{n}+1}}-
\left[\rho_{11}(0)-{\overline{n}\over
{2\overline{n}+1}}\right]\delta,
\end{eqnarray}
and $\delta={\exp}\{-\gamma(2\overline{n}+1)\Delta t\} \, (\Delta
t=t_i-t_{i-1}, i=1,2)$. As for the $n$ times histories, the
calculation for the elementary
probabilities  will be straightforward. For instance,
\begin{eqnarray}
p(1,1,\cdots 1)&=&{\rm Tr}\left(P_1K^{t_n}_{t_{n-1}}[P_1K^{t_{n-1}}_{t_{n-2}}[P_1\cdots
K^{t_2}_{t_1}[P_1K^{t_2}_{t_1}[\rho_0]P_1] \cdots P_1] P_1]\right)\nonumber\\
               &=&{1\over{(2\overline{n}+1)^{n-1}}}
\left[\overline{n}+(\overline{n}+1)\delta\right]^{n-1}\rho_{11}(t_1).
\end{eqnarray}
Similarly, one may calculate the transition probabilities, {\it etc}.

\section{Degree of Decoherence}
\label{sec:Degree}
Physically, one would not  expect the decoherence takes place
exactly. Therefore the investigation of the approximate decoherence is
of importance. In practical problems, one can, at best,  only expect that
probability sum rules are satisfied up to order $\epsilon$, for some
constant $\epsilon <1$. Namely,  the interference
terms do not have to be exactly zero, but small than probabilities by 
a factor of $\epsilon$.
One simple inequality which turns out to be very
useful to the study of the degree of decoherence is \cite{dowker,mce}: 
\begin{equation}
\label{inequality}
|D[\underline{\alpha}. \underline{\alpha}']|^2 
\le \epsilon^2 D[\underline{\alpha}.
\underline{\alpha}]D[\underline{\alpha}'. \underline{\alpha}']\>.
\end{equation}
We say that a system decoheres to order $\epsilon$ if the decoherence
functional satisfies (\ref{inequality}). As shown in
\cite{dowker},
such a condition implies that the most  probability sum
rules will then be satisfied to order $\epsilon$.  

Based on  this two-level model we will study the degree of
decoherence in some detail. To begin with, we establish 
the following trace inequality which is 
useful to our studies of the approximate decoherence.\\
{\bf Theorem:} Suppose that $M$ and $N$ are two $n\times n$ 
positive definite matrices. Let $P$ and
$Q$ be two $n\times n$ Hermitian matrices satisfying
\begin{equation}
\label{condition}
                        QP=PQ=0.
\end{equation}
Then  
\begin{equation}
\label{th}
|{\rm Tr} \left(MPNQ)\right|^2\leq\epsilon^2 {\rm
Tr}\left(MPNP\right){\rm Tr}\left(MQNQ\right)
\end{equation}
where $\epsilon =\min\{\epsilon^M,\epsilon^N\}$, 
here $\epsilon^M =(\lambda^{M}_{\max}-\lambda^{M}_{\min})
/(\lambda^{M}_{\max}+\lambda^{M}_{\min}), \epsilon^N =
(\lambda^{N}_{\max}-\lambda^{N}_{\min})/(\lambda^{N}_{\max}+
\lambda^{N}_{\rm min}), \lambda^{M}_{\max},\lambda^{M}_{\min}$  
and $\lambda^{N}_{\max},\lambda^{N}_{\min}$     are the maximal and the minimal eigenvalues of
$M$ and $N$,
respectively.

Remark: In fact, the condition that both $M$ and $ N$ are the positive
definite matrices could be
generalized to that  one is positive definite, say $M$, while
another $N$ is
positive semidefinite. In this case, $\epsilon=\epsilon^M$. It is 
hoped that the above theorem is also useful in some other cases. The
theorem is proved in the Appendix. 

For a general initial state represented by $\rho_0$ (pure
or mixed state), the decoherence functional of two time points may be
written as
\begin{equation}
\label{decoherence3}
D[\underline{\alpha},\underline{\alpha}']={\rm Tr}\left
(P_{\pm}K^{t_2}_{t_1}\left [P_- K^{t_1}_0\left
[\rho_0\right] P_+\right ]\right).
\end{equation}
We now write
\begin{eqnarray}
 A&=&K^{t_1}_{0}[\rho_0],\label{not1}\\
 B&=&\tilde K^{t_2}_{t_1}[P_{\pm}].\label{not2}
\end{eqnarray}
Then Eq. (\ref{decoherence3}) may
be rewritten, in the new notation, as
\begin{equation}
\label{de1}
D[\underline{\alpha},\underline{\alpha}']= {\rm Tr}\left(BP_-AP_+\right).
\end{equation}
Note that
$\tilde K$
in (\ref{not2}) is the super-propagator for the projection operators
\begin{equation}
P(t)=\tilde K^t_0[P(0)].
\end{equation}
The evolution equation for the projection operators is given by
\begin{eqnarray}
\label{projection}
\dot P&=&{i\over \hbar}[H,P]\nonumber \\
      & &+{\gamma\over 2}(\overline {n}+1) (2a^{\dag}Pa 
- a^{\dag} aP -Pa^{\dag} a)\nonumber \\
      & &+{\gamma \over 2}\overline {n}(2aPa^{\dag}
 - aa^{\dag} P- aa^{\dag} P) ,
\end{eqnarray}
where $H, a, a^{\dag}$ are defined as before. Note that the evolution
equation for 
the projection operator $P$ is
different from that for the density operator $\rho$ (\ref{master}). This reflects the
difference between the Schr\"{o}dinger  and Heisenberg
pictures in the density operator formalism. The explicit form of Eq. (\ref{projection}) may be written 
\begin{eqnarray}
\label{solution2}
\dot P_{11}&=&\gamma(\overline {n}+1)(P_{22}-P_{11})\>,\\
\dot P_{22} &=& \gamma\overline {n}(P_{11}-P_{22})\>,\\
\dot P_{12} &=&-[{\gamma\over 2}(2\overline {n} +1) -i\omega]P_{12}\>,\\
\dot P_{21} &=&-[{\gamma\over 2}(2\overline {n}+1) +i\omega] P_{21}\>.
\end{eqnarray}
The general solutions to the above equations are
\begin{eqnarray}
\label{solution3}
P_{11}(t)&=&C_1 + C_2e^{-\gamma(2\overline{n} +1)t}\>,\\
P_{22}(t)&=&C_1 -{\overline {n}\over \overline{n} +1}C_2 e^{-\gamma(2\overline n +1)t}\>,\\
P_{12}(t)&=&C_3e^{-[{\gamma\over 2}(2\overline{n} +1)-i\omega] t}\>,\\
P_{21}(t)&=&C_4e^{-[{\gamma\over 2}(2\overline{n} +1)+i\omega]t}\label{so9}\>,
\end{eqnarray}
where $C_i\> (i=1,2,3,4)$ are arbitrary constants. For given initial values,
these constants  can be expressed as 
\begin{eqnarray}
\label{solution4}
C_1 &=&{\overline{n}\over 2\overline{n}+1}P_{11}(0)+
            {\overline{n}+1\over 2\overline{n}+1}P_{22}(0)\>,\\ 
\label{solution5}
C_2&=&{\overline{n}+1\over 2\overline{n}+1}(P_{11}(0)- P_{22}(0))\>,\\
\label{solution6}
C_3&=&P_{12}(0)\>,\\
\label{solution7}
C_4&=&P_{21}(0)\>.
\end{eqnarray}

From the definitions (\ref{not1}) and (\ref{not2}), it is easy to see 
that in general, both
$A$ and $B$ could be 
positive definite matrices, and since $P_-$ and $P_+$ are projection operators, so the
condition (\ref{condition}) is automatically satisfied. Using the theorem
above, we immediately arrive at
\begin{equation}
|{\rm Tr}\left(BP_+AP_-\right)|^2 \leq\epsilon^2{\rm Tr}
\left(BP_+AP_+\right){\rm Tr}\left(BP_-AP_-\right)\>.
\end{equation}
That is,
\begin{equation} 
\label{mixted}
|D[\underline{\alpha},\underline{\alpha}']|^2\leq 
\epsilon^2 D[\underline{\alpha}, \underline{\alpha}])D[\underline{\alpha}',\underline{\alpha}']\>.
\end{equation} 
where $\epsilon ={\rm
min}\{\epsilon^A, \epsilon^B\}$, 
\begin{eqnarray}
\label{e1}
\epsilon^A&=&|\lambda^A_1-\lambda^A_2|\>,\\
\epsilon^B&=&{|\lambda^B_1-\lambda^B_2|\over\lambda^B_1+
\lambda^B_2}\label{e2}\>.
\end{eqnarray}
Here $\lambda^A_i\, (i=1,2)$ and $\lambda^B_i \,(i=1,2)$ are
two eigenvalues of $A$ and $B$, respectively. (Note that
$\lambda^A_1+\lambda^A_2=1$) From Eqs. (\ref{e1}) and (\ref{e2}), 
it is easily seen that the degree of
decoherence may  depend on  both the  projection operators we use
and the initial state of the system. This is also an expected
result. For the two-level system,  $\epsilon^A$
and $\epsilon^B$ can be calculated exactly. Consider, first, the
eigenvalues of $A$. Since Eq. (\ref{master}) preserves the trace, 
$\epsilon^A$ can be written as
\begin{equation}
\epsilon^A =\sqrt{1-4\lambda^A_1\lambda^A_2}=\sqrt{1-4\det A}\>.
\end{equation}
The determinant of $A$ can be explicitly evaluated from the general 
solutions (\ref{so}, \ref{so1}, \ref{so2}, \ref{so3}),
\begin{eqnarray}
\label{deta}
\det A &=&{\overline{n}(\overline{n}+1)\over (2\overline{n}+1)^2}+
              \left[{\overline{n}\over 2\overline{n}+1}\rho_{22}(0)+
               {\overline{n}+1\over 2\overline{n}+1}\rho_{11}(0)-
               {2\overline{n}(\overline{n}+1)\over(2\overline{n}+1)^2}
-\rho_{12}(0)\rho_{21}(0)\right]\delta\nonumber \\
           & &+\left[\rho_{11}(0)-{\overline{n}\over
2\overline{n}+1}\right]\left[\rho_{22}(0)-{\overline{n}+1\over
2\overline{n}+1}\right]\delta^2,
\end{eqnarray}
where $\delta =\exp\{-\gamma(2\overline{n}+1)t_1\}$. 
In order that a set of histories are decoherent, one expects
that $\delta$ should be small.

Similarly, $\epsilon^B$ can be
expressed as
\begin{equation}
\epsilon^B = \sqrt{1-{4\det B\over ({\rm Tr}B)^2}}\>.
\end{equation}
From (\ref{solution3})--(\ref{so9}), ${\rm Tr}B$ and
$\det B$ can be  easily obtained:
\begin{eqnarray}
{\rm Tr}B &=& 2C_1 +{1\over\overline{n}+1}C_2\delta_1,\\ 
\det B&=& C_1^2+\left[{1\over
\overline{n}+1}C_1C_2-C_3C_4\right]\delta_1-
{\overline{n}\over \overline{n}+1}C_2^2\delta_1^2,
\end{eqnarray}
where $\delta_1=\exp\{-\gamma(2\overline{n}+1)(t_2-t_1)\}$. 
The above discussions   show explicitly how  the degree of
decoherence is related to  the projection operators, the initial states
and the temperature of bath, as well as the time spacing
interval, in accordance with our general expectations.

In the long time limit, the density matrix will tend to the stationary
density matrix. Then we may get a much simpler expression for $\epsilon^A$:
\begin{equation}
\epsilon^A\sim {1\over 2\overline{n}+1}.
\end{equation}
As mentioned before, for the decoherent histories, 
$\delta$ and $\delta_1$ should be small. If we only keep the terms 
up to the first order of $\delta_1$, then $\epsilon^B$ becomes  
\begin{equation}
\epsilon^B\sim \left[{C_3C_4\delta_1\over {C_1^2 + {1\over \overline{n}+1}
C_1C_2\delta_1 }}\right]^{1/2}.
\end{equation}
The expression for $\epsilon^A$ can be obtained from Eq. (\ref{deta}).

It is seen from the above expressions that the degree of 
decoherence improves as the bath temperature increases.  We also
see that the projections with the smaller off-diagonal elements 
will give a better degree of decoherence. For a given system
with the initial state, then the matter for investigation is to
determine which histories, i.e., which string of projections, will
lead to the decoherence condition being satisfied. Therefore, we see
that $\epsilon^B$ serves as the main criterion for the degree of 
decoherence.

It is also of interest to compute the von Neumann entropy
 of $\rho(t)$\cite{zur2,zur3}. We
find that the von Neumann entropy supplies a  restriction on  the degree
of the decoherence of histories. It   exhibits a  tension  between
the predictability of the quantum  state  and the
degree of decoherence.

The von Neumann entropy  provides a
 convenient measure of the loss of predictability:
\begin{equation}
\label{entropy}
S=-{\rm Tr}\left(\rho\ln\rho\right),
\end{equation}
By definition, the more predictable state (pure state) may have 
less increase of the
entropy in  a fixed time period. This characterization process of
predictability is called
the  ``predictability sieve'' (coined by Zurek) which has been studied
recently in quantum Brownian motion model by using the linear entropy
\cite{zur2,zur3,gal}. We will see that two-level system provides a
very simple model to exhibit the relationship between the degree of
decoherence and entropy  by directly using the von Neumann entropy. 

For the purpose of the evaluating the entropy, we choose a special
basis in which  $\rho$ is diagonal. Let $\lambda_1$ and $\lambda_2$
be the eigenvalues of $\rho$, then Eq. (\ref{entropy}) reduces to
\begin{equation}
\label{entropy1}
S=-\sum_{i=1}^2\lambda_i\ln \lambda_i\>.
\end{equation}
Obviously, $\lambda_1$ and $\lambda_2$ can be expressed as
\begin{equation}
\lambda_1={1+\epsilon^A\over 2},\,\,\lambda_2={1-\epsilon^A\over 2}\>.
\end{equation}
Hence, Eq. (\ref{entropy1}) can be rewritten as
\begin{equation}
S=-\left[{(1+\epsilon^A)\over 2}\ln{(1+\epsilon^A)\over 2} +
{(1-\epsilon^A)\over 2}\ln{(1-\epsilon^A)\over 2}\right].
\end{equation}
It can be  shown  that $S(\epsilon^A)$ is a monotonically decreasing
function of $\epsilon^A$. Here, we find that there is a tension
between the predictability and the degree of decoherence. That is, the
initial density matrix which leads to less entropy production will
give worse degree of decoherence, namely, $\epsilon^A$ would be not
small (Here we are not considering $\epsilon^B$). 
This tension between the predictability and  the degree of the 
decoherence is a physically expected result. To obtain the higher
degree of the decoherence one would expect that the environment 
has stronger influence on the system of interest, such as 
increasing the temperature of the bath. Then the predictability
of the state, correspondingly, decreases.   

As an example, it is easy to see that the pure states 
$|1\ra$ and $|2\ra$ will lead to the largest entropy production.
However, we have seen that the histories consisting of the projection 
onto these two levels give the best degree of decoherence (In this case $\epsilon^B$
is zero). In fact, it has been shown that
the degree of decoherence is related to both the initial 
density matrix and projection operators used in the histories.

\section{unravelling of master equation}
\label{se:unravlling}

The master equation  provides an ensemble description of a
quantum system. The unravelling of master equation as  the stochastic
Schr\"{o}dinger equation for the state vector has provided many insights
into the foundation of quantum theory, especially in quantum
measurement  and the  useful tools to study
various practical problems in the quantum optics ({\it e.g.}, see
\cite{gis1,gar1}). In this section we
will study the localization in the  two different unravellings of the master equation
-- quantum jump simulation and quantum state
diffusion approaches. The former use the discrete random variables
whereas the latter use the continuous random variables.  

\subsection{Quantum Jump Simulation}
\label{sub:qujump}

The stochastic Schr\"{o}dinger equation used in  the quantum jump simulation 
describes a single quantum  process. The  system undergoes a
smooth evolution  until a jump takes place. This jump process is characterised
by the  real random variables which only take the discrete values. For the
master equation (\ref{master}), the stochastic Schr\"{o}dinger equation takes the
following form:
\begin{eqnarray}
\label{st:stochastic}
|d\psi\rangle &=&-{i\over \hbar}H|\psi\rangle dt\nonumber\\
        & &+\sum_{i=1}^2\left({{L_i\over \sqrt{\langle N_i\rangle}}-
            1}\right)|\psi\rangle dW_i\nonumber\\
        & & +\sum_{i=1}^2\left({\langle N_i\rangle\over 2} -{N_i\over 2}\right)dt\>.
\end{eqnarray}
Here $L_1=\sqrt{\gamma(\overline{n}+1)}a, L_2=\sqrt{\gamma\overline{n}}a^{\dag}$ are the Lindblad operators representing the
influence of the environment and $N_i= L_i^{\dag}L_i\> (i=1,2)$.
$\langle N_i\rangle =\langle\psi |N_i|\psi\rangle$ represents quantum average and $M$ represents the ensemble average.  
The   real random variables $dW_i\> (i=1,2)$ satisfy
\begin{eqnarray}
\label{real}
dW_idW_j&=&\delta_{ij}dW_i\>,\\
 M(dW_i)&=&\langle N_i\rangle dt\>\>\>  (i=1,2)\label{real1} .
\end{eqnarray}
Under condition (\ref{real}), it is easy to see that  $dW_i$ only take two
values: $0$ and $1$. The master equation (\ref{master}) can be
recovered from the stochastic Schr\"{o}dinger equation (\ref{st:stochastic})
in the sense that if $|\psi\rangle$ is the solution to
Eq. (\ref{st:stochastic}) then $\rho = M|\psi\rangle\langle\psi|$
satisfies master equation (\ref{master}).

In what follows we shall discuss the the `localization' properties of
the single jump trajectories. Here, by `localization' we mean that the
quantum state vector generated by the stochastic Schr\"{o}dinger equation will
converge to some fixed states in the mean square . The physical
meanings of this localization will become  clear from the later
discussions. 

More precisely, let $A$ be an
operator (not necessarily Hermitian), then we define the quantum mean square
deviation as
\begin{equation}
\label{qsd:deviation}
\sigma(A,A)=\langle A^{\dag}A\rangle - \langle
A^{\dag}\rangle \langle A\rangle .
\end{equation}
If the solution of the stochastic Schr\"{o}dinger equation
(\ref{st:stochastic}) satisfies
\begin{equation}
\label{co:conv}
M{d\over dt}\sigma(A,A)\leq 0\>,
\end{equation}
namely, the dispersion of the operator $A$ tends to decrease as time
evolves. Then we say that  the solution localizes at the eigenstates of the
operator $A$ ($A$ is sometimes called the collapse operator).

For the stochastic Schr\"{o}dinger equation for the quantum jump
simulation in two-level system, the collapse operator is
$\sigma_z$. Then quantum mean square deviation in this case is
\begin{equation}
\label{deviation}
(\Delta \sigma_z)^2=1-\langle \sigma_z\rangle^2\>.
\end{equation} 
In order to prove the localization, we should first derive  the
evolution equation of the expectation value of $\sigma_z$  by using the following
formula:
\begin{equation}
d\langle A\rangle =\langle \psi |A|d\psi\rangle +\langle d\psi |A|\psi\rangle+\langle d\psi |A|d\psi\rangle\>,
\end{equation} 
where $A$ is an operator. From (\ref{st:stochastic}), it is straightforward to arrive at the
following equation
\begin{eqnarray}
\label{evolution}
d\langle \sigma_z\rangle &=&(1-\langle \sigma_z\rangle)dW_1 -
                              (1+\langle \sigma_z\rangle)dW_2\nonumber\\
                         & & +\left[\langle \sigma_z\rangle\langle
                               N_1+N_2\rangle + \langle N_1-N_2\rangle\right]dt\>.
\end{eqnarray} 
Notice that 
\begin{equation}
d(\Delta \sigma_z)^2 = -2\langle \sigma_z\rangle d\langle
\sigma_z\rangle - (d\langle \sigma_z\rangle)^2\>.
\end{equation}
Then, inserting  Eq. (\ref{evolution}) into the above equation,
taking the ensemble means and remembering (\ref{real1}), we
obtain
\begin{eqnarray}
\label{means}
M{d\over dt}(\Delta \sigma_z)^2 &=&
-{1\over 2}\gamma(\overline{n}+1)(1-\langle\sigma_z\rangle)^2(1+\langle\sigma_z\rangle)\nonumber\\
                                & &-{1\over
2}\gamma\overline{n}(1+\langle\sigma_z\rangle)^2(1-\langle\sigma_z\rangle)\>.
\end{eqnarray}
The right-hand side of Eq. (\ref{means}) is
non-positive, and that it vanishes if and only if
$|\psi\rangle$ is $|2\rangle$ or $ |1\rangle$. Hence we conclude that
the solution to the stochastic Schr\"{o}dinger equation (\ref{st:stochastic})
will localize at $|2\rangle$ or $|1\rangle$ after a certain time. That is,
any initial state (which will be a superposition of $|1\ra$ and
$|2\ra$) will tend to a solution in which the atom undergoes
stochastic jumps between $|1\ra$ and $|2\ra$.

Let us now estimate this localization time. From
(\ref{means}), a few manipulations  directly give
\begin{equation}
\label{est}
M{d\over dt}(\Delta \sigma_z)^2 \leq -\gamma(2\overline{n} +1)(\Delta \sigma_z)^2 \>.
\end{equation}
So the localization rate $t_{\rm localization}$ is
\begin{equation}
\label{rate}
t_{\rm localization}\sim {1\over {\gamma(2\overline{n} +1)}}\>,
\end{equation}
which agrees with the decoherence time scale
(\ref{scale}).

In order to see the meanings of the {\it localization} of the quantum jump
process, let us compute the evolution of the populations of the two levels:
\begin{eqnarray}
\label{po1}
d|\langle 1|\psi\rangle |^2 &=&{1\over 2}(\langle
\sigma_z\rangle -1)dW_1 +
                                        {1\over 2}(1+\langle \sigma_z\rangle)dW_2\nonumber\\
                                        & & -{1\over 2}\left[\langle \sigma_z\rangle\langle
                                        N_1+N_2\rangle + 
                                        \langle N_1-N_2\rangle\right]dt\>,\\
 d|\langle 2|\psi\rangle |^2&=&{1\over 2}(1-\langle \sigma_z\rangle)dW_1 -
                              {1\over 2}(1+\langle \sigma_z\rangle)dW_2\nonumber\\
                         & & +{1\over 2}\left[\langle \sigma_z\rangle\langle
                               N_1+N_2\rangle + \langle  N_1-N_2\rangle\right]dt\>.
\end{eqnarray}  
Then it follows that from the above equations,
\begin{eqnarray}
|\langle 1|\psi\rangle|^2 &\longrightarrow & {\overline{n}+1\over {2\overline{n}+1}}\>,\\  
|\langle 2|\psi\rangle|^2&\longrightarrow & {\overline{n}\over {2\overline{n}+1}} \>.
\end{eqnarray}
That is to say, due to the influence of the environment to the system,
after a localization time scale (\ref{rate}), the average
populations of the first and second levels will become constant. This
may assist to understand the meanings  of the {\it localization} in the quantum
jump simulation.

In some sense, that the localization in quantum jump simulation  
chooses the basis $|1\ra, |2\ra$ appears to be natural, since
they correspond to the trajectories that would actually
observed in an individual experiment. As expected, the set of
histories consisting of projection onto the basis 
give the best degree of decoherence. In addition, we have
seen that density matrix become diagonal in this basis. 
Here, we have demonstrated a close connection between the different
approaches.

\subsection{Quantum state diffusion }

\label{sec:quan}

In this subsection, we will illustrate the localization process in another unravelling of the master
equation -- the quantum state diffusion approach, which was  introduced by Gisin and
Percival\cite{gis}  to describe the quantum open system by using
a stochastic Schr\"{o}dinger equation ( which is often called the
Langevin-Ito  
stochastic differential equation)  for the normalized
pure state vector of an individual system
of the ensemble. Similar to the quantum jump simulation, a solution
of the Langevin-Ito equation for the diffusion of
a pure quantum state in state space represents a single member of an
ensemble whose density operator satisfies the corresponding master
equation.

Generally, if the master equation takes the standard Lindblad form:
\begin{equation}
\label{mst}
\dot\rho=-{i\over \hbar}[H,\rho]+\sum_i(L_i\rho L^{\dag}_i-{1\over
2}L^{\dag}_iL_i\rho -{1\over 2}\rho L^{\dag}_iL_i)
\end{equation}
Then, correspondingly, the Langevin-Ito stochastic  equation can be written as 

\begin{eqnarray}
\label{langevin}
|d\psi\ra &=&-{i\over \hbar}H|\psi\ra dt\nonumber\\
        & &+\sum_i(\la L^{\dag}_i\ra L_i-{1\over
             2}L^{\dag}_iL_i-{1\over 2}\la L^{\dag}_i\ra\la
L_i\ra)|\psi\ra dt\nonumber\\
        & & +\sum_i(L_i-\la L_i\ra)|\psi\ra d\xi_i,
\end{eqnarray}
where $H$ is a Hamiltonian (of the open system) and $L_i$ are Lindblad
operators, as before, $\la L_i\ra=\la\psi| L_i|\psi\ra$. The complex Wiener processes 
$ d\xi_i$ satisfy 
\begin{equation}
 M(d\xi_i)=0, M(d\xi_id\xi_j)=0, 
M(d\xi_i^{\ast}d\xi_j)=\delta_{ij}dt,
\end{equation}
where $M$ denotes a mean over the ensemble. Quantum state diffusion
reproduces the master equation in the mean:
\begin{equation}
\rho= M|\psi\ra\la\psi|,
\end{equation}
where $|\psi\ra$ satisfy the quantum state diffusion equation (\ref{langevin}), then it can
be shown that $ \rho$ satisfies the master equation (\ref{mst}).

In order to show the localization properties of the Langevin-Ito
equation we now consider the simplest case which is assumed that bath temperature 
is zero ($\bar n = 0$). In this case the master equation
(\ref{master})
reduces to
\begin{eqnarray}
\label{zero}
\dot \rho  = &-&{i\over \hbar} [H, \rho ]\nonumber \\
             &+& {\gamma\over 2} (2a\rho a^{\dag} - a^{\dag} a\rho -\rho a^{\dag} a),
\end{eqnarray}
Then the corresponding Langevin-Ito  equation is given by

\begin{eqnarray}
\label{sto}
|d\psi\ra &=&-{i\over \hbar}H|\psi\ra dt\nonumber\\
        & &+{\gamma\over 2}(2\la a^{\dag}\ra a-
             a^{\dag}a-\la a^{\dag}\ra\la a\ra)|\psi\ra dt\nonumber\\
        & & +\sqrt{\gamma}(a-\la a\ra)|\psi\ra d\xi,
\end{eqnarray}
where $d\xi$ is the complex Wiener process
satisfying 
\begin{equation}
M(d\xi )=0,\> M(d\xi d\xi )=0,\> M(d\xi^{\ast}d\xi )=dt,
\end{equation}
where $M$ denotes a mean over probability distribution.

The evolution of the quantum average of operators can be calculated by using the
following formula: 
\begin{eqnarray}
\label{exp}
d\la G\ra&=&{i\over\hbar}\la[H,G]\ra\nonumber \\
    & & -{1\over
         2}\sum_i\la L^{\dag}_i[L_i,G]+[G,L^{\dag}]L_i\ra dt\nonumber\\ 
    & & + \sum_i(\sigma(G^{\dag},L_i)d\xi_i+\sigma(L_i,G)d\xi_i^{\ast}),
\end{eqnarray}
where
\begin{equation}
\sigma (A,B)=\langle A^{\dag}B\rangle -\langle A^{\dag}\rangle \langle B\rangle
\end{equation}
Using Eq. (\ref{exp}), it is straightforward to get the following equations: 
\begin{eqnarray}
d\langle\sigma_x\rangle &=&\left [-{\omega\over \hbar}\la\sigma_y\ra-{\gamma^2\over
               2}\la\sigma_x\ra\right ]dt\nonumber \\
            & &+{\sqrt {\gamma}\over 2 }\left [1+\la\sigma_z\ra-\la\sigma_x\ra^2
               +i\la\sigma_x\ra\la\sigma_y\ra\right ]d\xi\nonumber \\
            & &+{\sqrt {\gamma}\over 2} \left [1+\la\sigma_z\ra-\la\sigma_x\ra^2
               -i\la\sigma_x\ra\la\sigma_y\ra\right ]d\xi^{\ast},\\
d\la\sigma_y\ra &=&\left [-{\omega\over \hbar}\la\sigma_x\ra-{\gamma^2\over
               2}\la\sigma_y\ra\right ]dt\nonumber \\
            & &+{\sqrt {\gamma}\over 2 }\left [-i(1+\la\sigma_z\ra)
+i\la\sigma_y\ra^2
               -\la\sigma_x\ra\la\sigma_y\ra\right ]d\xi\nonumber \\
            & &+{\sqrt {\gamma}\over 2} \left [i(1+\la\sigma_z\ra)
-i\la\sigma_y\ra^2
               -\la\sigma_x\ra\la\sigma_y\ra\right ]d\xi^{\ast}\\
d\la\sigma_z\ra&=&-[\la\sigma_z\ra\gamma+\gamma]dt\nonumber\\
           & &-{\sqrt {\gamma}\over 2}(1+\la\sigma_z\ra)(\la\sigma_x\ra-
               i\la\sigma_y\ra)d\xi\nonumber\\
           & &-{\sqrt {\gamma}\over 2}(1+\la\sigma_z\ra)(\la\sigma_x\ra+i\la\sigma_y\ra)d\xi^{\ast}.
\end{eqnarray}
Moreover, we need to calculate the higher order moments. For any
Hermitian operator $A$ we have from (\ref{qsd:deviation}), 
\begin{eqnarray}
d(\Delta A)^2&=&d(\la A^2\ra-\la A\ra^2)\nonumber\\
             &=&d\la A^2\ra-2\la A\ra d\la A\ra-(d\la A\ra)^2.
\end{eqnarray}
Then we easily obtain
\begin{eqnarray}
M{d\over at}(\Delta \sigma_x)^2&=&2\omega
                                       \la\sigma_x\ra\la\sigma_y\ra+\gamma \la\sigma_x \ra^2\nonumber \\
                                    & & -{\gamma\over 2}(\Delta
                                        \sigma_x)^4-\gamma(\Delta\sigma_x)^2\la\sigma_z\ra\nonumber \\
                                    & & -{\gamma\over 2}\la\sigma_z\ra^2-{\gamma\over 2}(\la\sigma_x\ra\la\sigma_y\ra)^2,\\
M{d\over at}(\Delta \sigma_y)^2&=&-2\omega
                                       \la\sigma_x\ra\la\sigma_y\ra+\gamma \la\sigma_y \ra^2\nonumber \\
                                    & & -{\gamma\over 2}(\Delta
                                        \sigma_y)^4-\gamma(\Delta\sigma_y)^2\la\sigma_z\ra\nonumber\\
                                    & &-{\gamma\over 2}\la\sigma_z\ra^2-{\gamma\over 2}(\la\sigma_x\ra\la\sigma_y\ra)^2.
\end{eqnarray}
Now, we are in the position to consider the localization of solutions to
Eq. (\ref{sto}). Using master equation or quantum trajectories approach, it is
very easy to see that the atom will soon collapse into the lower state
$|1\ra$ and keeps there forever. Here we shall demonstrate that any solution
to Langevin-Ito  equation (\ref{sto}) will localize at the lower state after a
localization time. The collapse operator in this case is
\begin{equation}
A=\sigma_x+i\sigma_y.
\end{equation}
Then by using (\ref{qsd:deviation}) we get
\begin{equation}
\sigma(A,A)=(\Delta \sigma_x)^2+(\Delta \sigma_y)^2+2\la\sigma_z\ra.
\end{equation}
Hence  we have
\begin{eqnarray}
\label{means1}
 M{d\over dt}\sigma
(A,A)&=&{\gamma}\la\sigma_x\ra^2+{\gamma}\la\sigma_y\ra^2-{\gamma\over 2}(\Delta
               \sigma_x)^4\nonumber\\
            & &-{\gamma\over 2}\gamma(\Delta \sigma_y)^4-\gamma(\Delta\sigma_x)^2\la\sigma_z\ra-
               \gamma(\Delta\sigma_y)^2\la\sigma_z\ra\nonumber\\
            & &-\gamma\la\sigma_z\ra^2-2\gamma(\la\sigma_z\ra+1) -\gamma(\la\sigma_x\ra\la\sigma_y\ra)^2.
\end{eqnarray}
In order to prove that the left-hand side of Eq. (\ref{means1}) is non-positive, let us denote
\begin{eqnarray}
(\Delta \sigma_x)^2&=&1+X\label{pa2},\\
(\Delta \sigma_y)^2&=&1+Y\label{pa3},\\
\la\sigma_z\ra         &=&-1+Z .\label{pa4}
\end{eqnarray}
Substituting equations (\ref{pa2}),(\ref{pa3}) and  (\ref{pa4}) into
Eq. (\ref{means1}) we have
\begin{equation}
\label{positive}
{\rm M}{d\over dt}\sigma (A^{\dag},A)={\gamma}
\left [-R^2-X-Y-2Z-{1\over 2}(Y-Z)^2-{1\over 2}(X-Z)^2\right ],
\end{equation}
where $R=\la\sigma_x\ra\la\sigma_y\ra$. Note that
\begin{equation}
\label{positive1}
X+Y+2Z =\sigma (A,A)\geq 0.
\end{equation}
Then we show that 
\begin{equation}
\label{means2}
 M{d\over dt}\sigma (A^{\dag},A)\leq 0
\end{equation}
and the equality holds if and only if
\begin{equation}
X=Y=Z=0.
\end{equation}
That is,  the average in the left hand sides of equations (\ref{pa2}),
(\ref{pa3}), and ({\ref{pa4}) is taken over the ground state
$|1\ra$. This proves that the solution  to Eq. (\ref{sto}) will localize at
the ground state when the evolution time is
larger than the  localization time.

Finally, let us estimate the localization rate of the quantum state evolution. 
Using Eq. (\ref{positive}) and Eq. (\ref{positive1}), 
we immediately obtain

\begin{equation}
M{d\over dt}\sigma (A,A)\leq -\gamma (\sigma (A,A))^2.
\end{equation}
So the localization rate $t_{\rm localization}$ is
\begin{equation}
t_{\rm localization} \sim {1\over \gamma}.
\end{equation}
Again we see that the above result is 
in agreement with the decoherence time scale (\ref{scale}).

\section{Concluding remarks}
In this paper, based on the two-level system models,
we have studied in detail the decoherent histories 
approaches. We present an explicit analysis of the 
degree of decoherence  and its relation to the von
Neumann entropy. We have demonstrated the localization 
in both quantum jump simulations and quantum state 
diffusion approaches. Here we conclude with a few 
remarks.

Firstly, we have shown that the environment-induced
decoherence, decoherent histories and the  localization 
process in the stochastic Schr\"{o}dinger equation 
approaches generally agree each other. We have shown 
that the set of  histories consisting of the projection onto
the basis in which the density operator tends to become
diagonal give the better degree of decoherence. These 
results are in agreement with former
studies on the quantum Brownian models
\cite{hal} as well as the quantum optical models \cite{twa}.

In this paper, we have shown that there are a number 
of sets of decoherent histories in this two levels model.
Clearly, these decoherent histories are not
equally important from physical point of view. Among 
those, the most natural one is that which consist of the
projections onto $|1\ra$ and $|2\ra$. We have proven that 
this set of histories give the best degree of decoherence.
Note that the density matrix in the basis $|1\ra$ and 
$|2\ra$ will become diagonal after a typically short time. 
Moreover, we show that the solutions to the stochastic 
Schr\"{o}dinger equation in the quantum jump simulation
will localize at $|1\ra$ or $|2\ra$ after certain
time which is basically same as the decoherence time.

It is known that the solution generated by stochastic
Schr\"{o}dinger equation (\ref{st:stochastic}) will randomly 
jump between the two levels $|1\ra$ and $|2\ra$. However, due to 
the influence of the bath, the average populations of
the two levels will gradually become stable, and the localization 
process occurs. As we have shown, the set of histories 
consisting of the projection onto these two levels are
perfectly decoherent. Here, we have seen that quantum 
jump simulation is entirely compatible with 
the history point of view. This is a very nice result. 
Similar results in  the quantum state diffusion have been 
discussed before \cite{dio1,hal}

In addition, we have found that the environment-induced 
decoherence, decoherent histories and the localization 
process are more effective as the bath temperature 
increases. Physically, this is an expected result as the
bath at a higher temperature would have stronger influence
on the system.  As expected, the time scales concerning 
both decoherence and localization are basically same.

Secondly, the approximate decoherence is of basic importance in
practical physical process.  By using this two-level system model
we can clearly see what determines the degree of decoherence. For
a given set of histories, the only adjustable parameters are the
temperature of bath, the time-spacing interval and the initial state
of the system. We also see a interesting relation between the 
degree of decoherence and the von Neumann entropy. This relation
indicates that there is a tension between the predictability of state
of system and the degree of decoherence. A similar tension has
been discussed in a different situation based on quantum Brownian 
motion models \cite{gel1}.

Thirdly, it is important to notice that, as phenomenological theories, 
both quantum state diffusion and quantum  jump simulation must
be used under some conditions (e.g. see\cite{mil}). The comparison
between different approaches therefore  must be made in caution 
since the correspondence between  them is by no means 
mathematically one-to-one correspondence. Rather, we emphasize 
that, underlying the open quantum system,  the mutual influence
between the system and its  environment 
is the common theoretical base of all of those approaches and  both 
decoherence and localization are nothing more than the different 
manifestations of a single
entity.

Finally, let us note that the coarse-graining in our two-level system 
is made by using projection operators on the system whilst ignoring the
environment. It would be interesting to consider the general
$n$-dimensional model in which the effect of a further coarse-graining
on the degree of decoherence can be discussed. Work towards 
to this aspect is in progress. 

The environment-induced decoherence, decoherent histories as
well as various stochastic Schr\"{o}dinger equations  have provided 
many important insights into the understanding of fundamental 
problems in quantum theory. The investigation into the similarity
and difference between the different approaches is of importance.
The more thorough studies in this aspect would be useful.

\section*{Acknowledgements}
The author would like to express his sincere thanks to J. J. Halliwell
for suggesting this project, for encouragements, and for many
suggestions which are critically important for the ideas in this 
paper. He is also grateful to A. Zoupas and B. Meister for stimulating 
discussions.

This work was supported by  the SBFSS scholarship from the British Council.

\appendix

\begin{section}{Proof of Theorem}
In this Appendix, we shall give a proof of Theorem in Section III.  
Since both $M$ and $N$ are positive definite matrices, therefore, one
of them, say, $N$
can be decomposed  as
\begin{equation}
N=S^{\dag}S,
\end{equation}
where $S$ is an $n\times n$ matrix. After an arrangement, the
right-hand side of Eq. (\ref{th}) becomes
\begin{equation}
\label{tr}
|{\rm Tr}\left(MPS^{\dag}SQ\right)|=|{\rm Tr}\left(SQMPS^{\dag}\right)|.
\end{equation}
Suppose $x_m$ are an orthonormal basis in $n$ dimensional space $V$.
Then
\begin{equation}
\label{tr1}
|{\rm Tr}\left(MPNQ\right)|=|\sum_m x_m^T\left(SQMPS^{\dag}\right)x_m|,
\end{equation} 
where $x_m^T$ is the transpose of $x_m$. Now, we set
\begin{eqnarray}
\label{no}
y_m&=&QS^{\dag}x_m,\\ 
z_m&=&PS^{\dag}x_m.
\end{eqnarray}
Then the trace in Eq. (\ref{tr1}) may be rewritten as
\begin{equation}
\label{tr2}
{\rm Tr}\left(MPNQ\right)=\sum_m y_m^TMz_m.
\end{equation}
Since $y_m, z_m$ are orthogonal vectors and $M$ is a  positive
definite matrix, then it is not difficult to arrive at 
the following inequality (see \cite{mei})
\begin{equation}
\label{tr4}
|\left(y_m^TMz_m\right)|\leq \epsilon^M \left(y_m^TMy_m\right)^{1/2}
\left(z_m^TMz_m\right)^{1/2},
\end{equation}
where $\epsilon^M=(\lambda^M_{\max}-\lambda^M_{\min})/
(\lambda^M_{\max}+\lambda^M_{\min})$, $\lambda^M_{\max}$ and
 $\lambda^M_{\min}$ are the largest and the smallest eigenvalues of
$M$, respectively.
Combining  (\ref{tr4}) with Cauchy's inequality, 
\begin{equation}
(\sum_m a_nb_n)^2\leq \sum_ma_n^2\sum_mb_n^2,
\end{equation}
then (\ref{tr1}) becomes
\begin{eqnarray}
|{\rm Tr}\left(MPNQ\right)|&=&|\sum_m 
                                          x_m^T\left(SQMPS^{\dag}\right)x_m|\nonumber\\
                                      &\leq& \sum_m |y_m^TMz_m|\nonumber\\ 
                                      &\leq&\epsilon^M (\sum_m
y_m^TMy_m)^{1/2}  
(\sum_m z_m^TMz_m)^{1/2}.
\end{eqnarray}
It is easy to identify that
\begin{eqnarray}
{\rm Tr}\left(MPNP\right)&=&\sum_m
y_m^TMy_m,\\  
{\rm Tr}\left(MQNQ\right)&=&\sum_m z_m^TMz_m.
\end{eqnarray}   
This proves that
\begin{equation}
|{\rm Tr}\left(MPNQ\right)|\leq
\epsilon^M\left[{\rm Tr}\left(MPNP\right)\right]^{1/2}\left[ {\rm Tr}\left(MQNQ\right)\right]^{1/2}.
\end{equation}
Since $M$ and $N$ are in the completely symmetric position, so the similar result
is true for $\epsilon^N$. Then it completes the proof of  the theorem. $\Box$
\end{section}

\end{document}